\documentclass[final]{IEEEtran}

\usepackage{graphicx}
\ifCLASSINFOpdf
\else
\fi
\PassOptionsToPackage{hyphens}{url}\usepackage{hyperref}

\usepackage{amsfonts,amsmath,amssymb}
\usepackage[ruled]{algorithm2e}
\usepackage[caption=false,font=scriptsize,labelfont=sf,textfont=sf]{subfig}
\usepackage{listings}
\usepackage{dsfont}
\usepackage{hyperref}
\usepackage{bm}

\newcommand{\R}{\mathbb{R}}

\begin{document}
%
\title{Adaptive Methods for Short-Term Electricity Load Forecasting During COVID-19 Lockdown in France}

%
%
%

\author{David Obst\textsuperscript{*}\thanks{* The first two authors have contributed equally to this work.}, Joseph de Vilmarest\textsuperscript{*}, Yannig Goude\thanks{D. Obst (david.obst@edf.fr), J. de Vilmarest (joseph.de-vilmarest@edf.fr) and Y. Goude (yannig.goude@edf.fr) are with \'Electricit\'e de France R\&D.}}

%
%

\markboth{}%
{Adaptive additive models for short-term electricity load forecasting during COVID-19 lockdown}
%



\maketitle

\begin{abstract} 
The coronavirus disease 2019 (COVID-19) pandemic has urged many governments in the world to enforce a strict lockdown where all nonessential businesses are closed and citizens are ordered to stay at home. One of the consequences of this policy is a significant change in electricity consumption patterns. Since load forecasting models rely on calendar or meteorological information and are trained on historical data, they fail to capture the significant break caused by the lockdown and have exhibited poor performances since the beginning of the pandemic. This makes the scheduling of the electricity production challenging, and has a high cost for both electricity producers and grid operators. In this paper we introduce adaptive generalized additive models using Kalman filters and fine-tuning to adjust to new electricity consumption patterns. Additionally, knowledge from the lockdown in Italy is transferred to anticipate the change of behavior in France. The proposed methods are applied to forecast the electricity demand during the French lockdown period, where they demonstrate their ability to significantly reduce prediction errors compared to traditional models. Finally expert aggregation is used to leverage the specificities of each predictions and enhance results even further.
\end{abstract}


%
\IEEEpeerreviewmaketitle

\section{Introduction}
%
%
%


\IEEEPARstart{A}{ccurate} electricity load forecasting is of paramount importance for the balancing of the electricity grid, since they are the main inputs of the production planning at different horizons \cite{bunn1985comparative} and storage capacities are still limited regarding the consumption needs. Load forecasting is performed at different horizons of time, ranging from intra-day (10 minutes to 24 hours ahead) to daily, weekly, monthly or even a few years in advance for industrial needs covering production planning, demand response, grid management, electricity trading, risk management, optimization of production units maintenance and commercialization. 

The field has been thoroughly studied the past decades, especially by the time series, statistics and machine learning communities. Time series approaches are very efficient for very-short term forecasts (typically less than 24 hours ahead). They rely on auto-regressive moving-average (ARIMA) models \cite{huang2003short} or functional approaches \cite{antoniadis2016prediction, cho2013modeling} exploiting daily and weekly patterns in the electricity load data. Statistical and machine learning models are usually stronger for short and mid-term predictions (more than 1 day  ahead). They use calendar characteristics (such as the time of the year, day of the week...) as well as meteorological effects (temperature, wind speed) or tariff options as inputs and are then trained on a large set of historical data (usually 3 to 5 years). A good overview of load forecasting practices has been given by the Global Energy Forecasting Competition (GEFCOM) 2012 \cite{hong2014global}. Popular algorithms include  gradient boosting machines \cite{lloyd2014gefcom2012}, neural networks \cite{park1991electric,ryu2017deep} or Generalized Additive Models (GAM) \cite{pierrot2011short, Goude2013, fasiolo2020fast}. These semi-parametric models are very attractive to electric utilities as they combine the flexibility of fully nonparametric models, the simplicity of multiple regression model and are computationally efficient to scale with big data \cite{wood2015generalized}. The main French electricity provider, EDF (\'{E}lectricité de France) uses GAM as their lead forecasting tool.


\begin{figure}[ht]
\centering
  \includegraphics[scale=0.5]{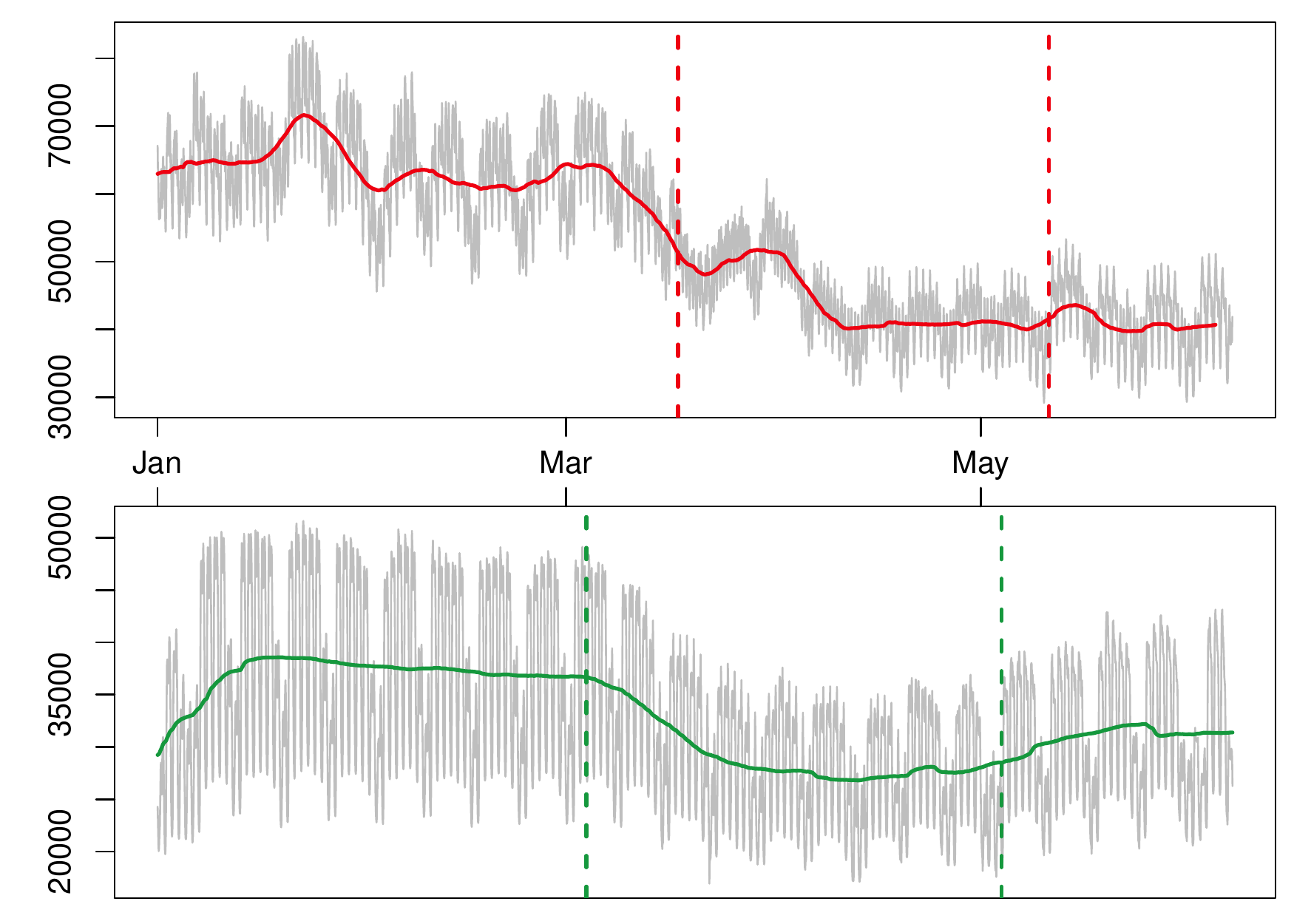}
  \caption{French and Italian electricity load (in MW) at resp. half-hourly and hourly resolution in 2020. Dashed lines are the starting and ending date of the lockdown.} \label{fig:evolution_covid}
\end{figure}

However the coronavirus pandemic has significantly affected consumption patterns all over the world. As presented in \cite{narajewski2020changes, IEA2020}, the closure of nonessential businesses as well as the stay-at-home directives have led to a significant drop of the power demand and changes in the daily consumption patterns. Figure \ref{fig:evolution_covid} shows the French and Italian electricity load over time in 2020, whose decrease due to the lockdown (which happens before in Italy) is clearly seen. Daily profiles of the French consumption before and after the lockdown are represented in Figure \ref{fig:profile_covid}. After lockdown for both countries the daily shapes of the load have converged towards the one of Saturdays.

\begin{figure}
\centering
\hspace{-0.8cm}
\begin{minipage}{.45\columnwidth}
  \centering
  \includegraphics[scale=0.25]{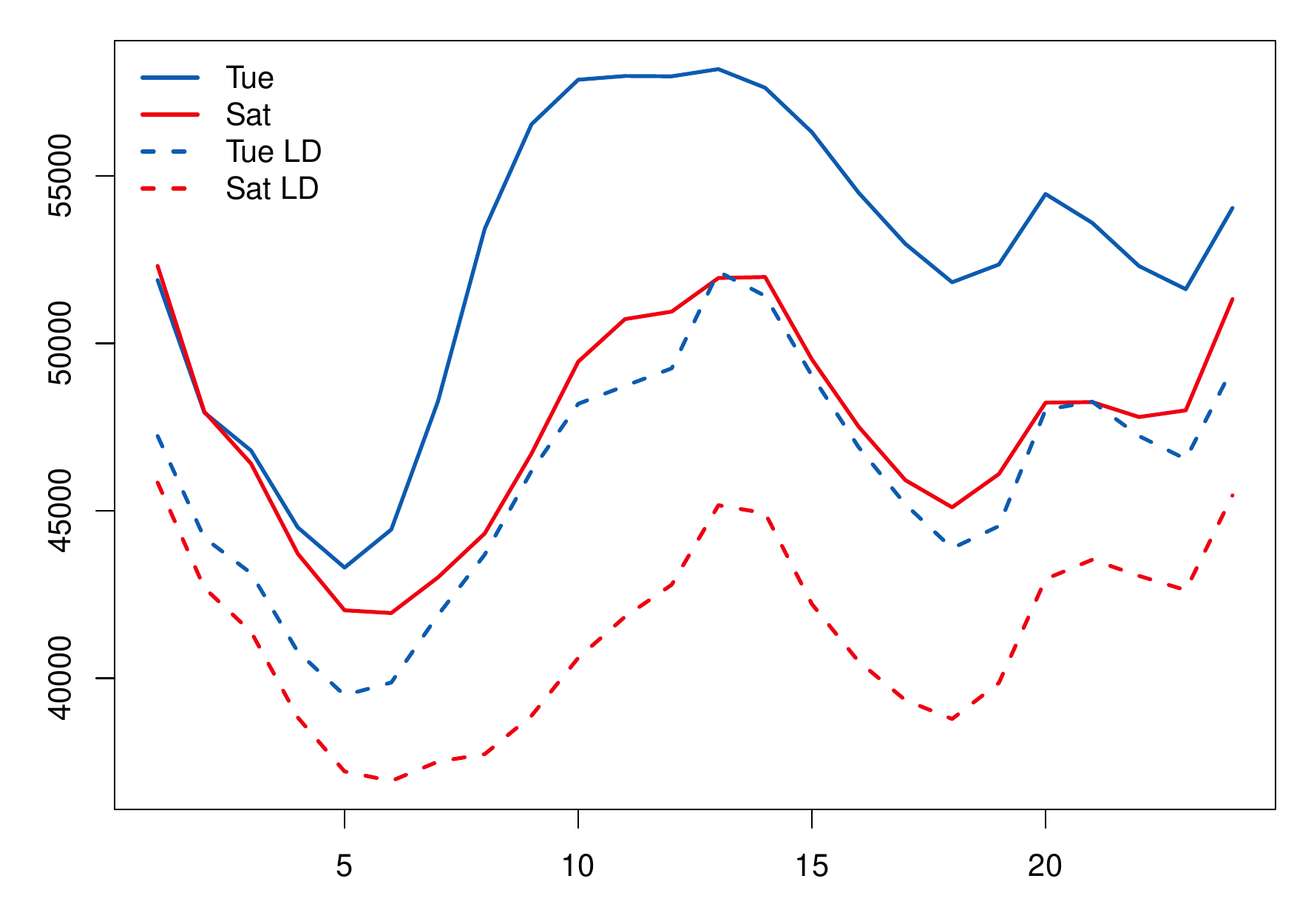} 
  \label{fig:test1}
\end{minipage}%
\hspace{0.4cm}
\begin{minipage}{.45\columnwidth}
  \centering
  \includegraphics[scale=0.25]{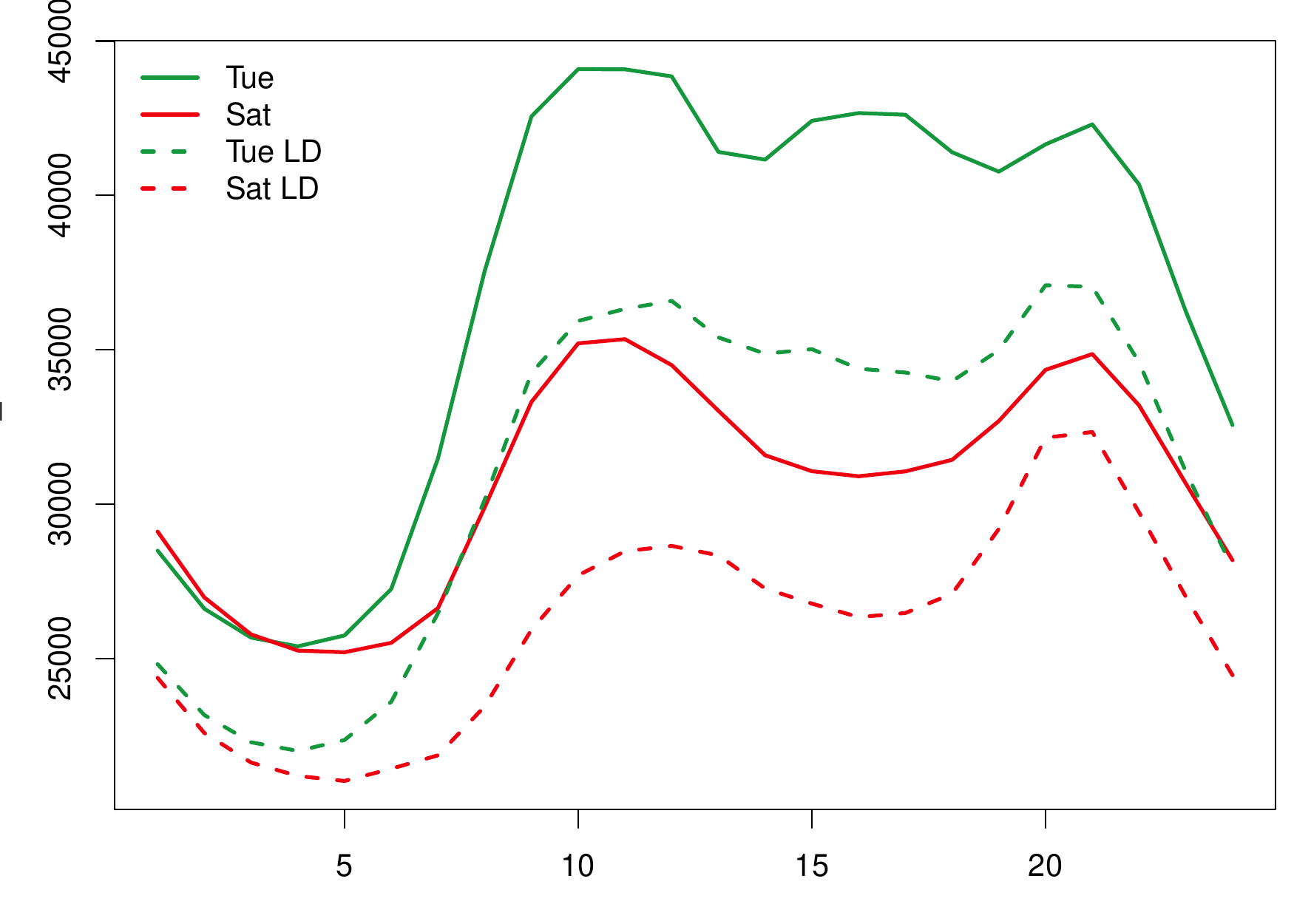}
  \label{fig:profile_covid}
\end{minipage}
\caption{French and Italian electricity Tuesday and Saturday load profiles before and during the lockdown (Dashed lines).} 
\label{fig:profile_covid}

\end{figure}

Since models are trained on historical data and make the underlying assumption that future behavior will be similar to past one, they will fail to produce satisfactory predictions during the lockdown period. For instance in France GAM usually achieve around 1\% MAPE (mean absolute percentage error) \cite{pierrot2011short}, but were around 5\% during the  first few weeks of the lockdown thus necessitating expert intervention to correct the model forecasts. Not only do these poor forecasts have a high cost for electricity producers and system operators, but they represent a threat to the proper functioning of the electrical network as well, which could have even more consequences than usual during a pandemic.

This is why finding novel approaches to better predict the load demand during these troubled times is of paramount importance. However to our knowledge, up to this date only a few papers have addressed this problem. \cite{nagbe2020covid} is among the first to propose an efficient strategy to improve the predictions during the COVID-19 lockdown period in France. Using an adaptive functional state-space model and assimilating the period to non-workable days, the author was able to achieve significantly better performance compared to the french system operator. In \cite{chen2020using} the integration of mobility data is combined with multi-task learning to improve the forecasting during the lockdown. They show that mobility is indeed a relevant feature that should be integrated in load demand models, and that joint training of a neural network for multiple geographical areas yields additional benefits and compensates for the lack of data. However none of these papers are investigating how GAM could be improved during the lockdown period. 

We consider here the framework of GAM and propose two new adaptive versions of these models. The idea of adaptive models is to take advantage of data observed in an online fashion to update an initial model. 
In every adaptive forecasting method a trade-off has to be found between a good reactivity to a change (whether it is a smooth drift or a break) and a good behavior during stable periods. One of the most popular algorithm for that is the Kalman filter \cite{kalman1960new} already applied to electricity load forecasting in \cite{harvey1993forecasting} and \cite{dordonnat2009dynamic}. We propose here to couple Kalman filters with GAM to obtain a forecasting procedure which performs well before the lockdown, exploiting the nice properties of GAM but also reacting quickly to the sudden change in the data at the beginning of the lockdown. 
The second approach we present leverages ideas from transfer learning to fine-tune a GAM on the lockdown period. Transfer learning (also  referred  as learning-to-learn or knowledge transfer) is a branch of machine learning that aims at reusing knowledge from one source task on another target one \cite{pan2010survey,weiss2016survey}. It has shown great success, particularly when the source data is plentily available and the target one scarce. Recently it has even found applications for electricity load forecasting to transfer information from one set of customers to another one \cite{cai2020forecasting}. In our case our source data will be the data before the lockdown and the target one the data during the lockdown in the country of interest (France in our study), or even a similar one where the lockdown came  before (e.g. Italy here). The contributions of our work are the following:

\begin{enumerate}
    \item Two mathematical approaches are proposed to efficiently adjust a historical model to consumer behavior change over time, even in the case where data is scarce. Furthermore they do not require the integration of additional features.
    \item The two methodologies have been successfully applied on the difficult period of the COVID-19 lockdown in France, achieving forecast accuracy close to the one observed before the pandemic.
    \item An empirical strategy is suggested to anticipate  the impact of the lockdown on the load using another country's data, thus enabling satisfactory predictions from the very first day of stay-at-home order.
\end{enumerate}

The rest of the paper is organized as following. In Section 2 we introduce the two model adaptation methods relying on Kalman filtering and fine-tuning. Section 3 presents the data and the GAM model used for the French load and Section 4 summarizes the main results of our experiments. Finally Section 5 concludes our study and suggests further work.


\section{Adaptation of additive models}
\label{section:gam_adapt}

We consider additive models whose assumption is that the response variable $y_t$ is decomposed as
\begin{equation*}
    y_t = \beta_0 + \sum_{j=1}^d f_j(x_{t,j}) + \varepsilon_t \,,
\end{equation*}
where $(\varepsilon_t)$ is an independent identically distributed (i.i.d.) random noise, $\bm{x}_t = (x_{t,1},...,x_{t,d})$ are the explanatory variables at time $t$, and each nonlinear effect $f_j$ is decomposed on a spline basis $(B_{j,k})$ with coefficients $\bm\beta_j$:

\begin{equation*}
    f_j(x) = \sum\limits_{k=1}^{m_j} \beta_{j,k} B_{j,k}(x) \,.
\end{equation*}
where $m_j$ depends on the dimension of the spline basis. The coefficients $\beta_0, \bm{\beta}_1, \dots, \bm{\beta}_{d}$ then are estimated by penalized least-squares. The penalty term involves the second derivatives of the functions $f_j$, forcing the effects to be smooth (see \cite{wood2017generalized}).

The random residuals $\varepsilon_t$ are supposed to be Gaussian i.i.d. in the first place. Later in the numerical experiments we will introduce another variant of this model, where the residuals are supposed to be an ARIMA model optimised with classical time series methods. We focus here on structural adaptation of the GAM over time. We present two different levels of adaptation. First, we consider the reduced problem of adapting a linear combination of the frozen effects $f_1,...,f_d$. Secondly we try to adapt the whole model by fine-tuning.

\subsection{Multiplicative correction of the effects}
In order to reduce the dimension of the adaptation problem, a strategy is to freeze the nonlinear effects, and to correct these effects by a multiplicative factor. Precisely, we define $f(\bm{x}_t) = (1,\overline{f}_1(x_{t,1}),...,\overline{f}_d(x_{t,d}))^\top$ where $\overline{f}_j$ is a normalized version of $f_j$ obtained by subtracting the mean on the train set and dividing by the standard deviation. Then we adaptively estimate $\bm\theta_t$ such that
\begin{equation*}
    \mathbb{E}[y_t] = \bm{\theta}_t^\top f(\bm{x}_t)\,.
\end{equation*}

Thus we reduce the number of coefficients from $1+\sum_{j=1}^d m_j$ to $1+d$. This is a good trade-off to obtain a simple model which will react quickly to a break in the data generation process but also complex enough to fit well with the nonlinear properties of the load.

\subsubsection{Exponential Least-Squares}
An empirical method consists in solving at each step a least-squares problem where we specify a weight decreasing exponentially with the time difference. Precisely we define
\begin{align*}
    & \hat{\bm{\theta}}_t = \arg\min\limits_{\bm{\theta}\in\mathbb{R}^d} \sum\limits_{s=1}^{t-1} e^{- \mu (t-s)} \Big(y_s-\bm{\theta}^\top f(\bm{x}_s)\Big)^2 \,,
\end{align*}
and we predict $\hat{y}_t = \hat{\bm\theta}_t^\top f(\bm{x}_t)$.
This formalisation leads to a single parameter, the exponential forgetting factor $\mu$. The advantage of this type of adaptation lies in its simplicity. The forgetting factor $\mu$ is determined by minimizing the RMSE on a validation set composed of the last year of the train set for a GAM trained on the beginning of the train set, then we keep the same $\mu$ for the GAM trained on the whole train set. Previous work has been done on estimating this parameter online, but leads to computational issues and potential instability of the model (see \cite{ba2012adaptive}).

\subsubsection{Kalman Filter}\label{section:kalmanadaptation}
We present also a state-space model approach. We assume the following equations:
\begin{align*}
    & y_t = \bm\theta_t^\top f(\bm{x}_t) + \varepsilon_t\,,\\
    & \bm\theta_{t+1} = \bm\theta_t + \bm\eta_t\,,
\end{align*}
where $(\varepsilon_t)$ and $(\bm\eta_t)$ are gaussian white noises of respective variance / covariance $\sigma^2$ and $Q$. This is the setting of Kalman filtering~\cite{kalman1960new}, thus we use the recursive formulae of Kalman providing the expectation and covariance of the state $\bm\theta_t$ given the past observations, and these estimators yield the mean and variance of $y_t$ given the past. This is described in Algorithm~\ref{alg:kalman}.

\begin{algorithm}[t]
{\caption{Kalman Filter}
\label{alg:kalman}}
{
{\bf Initialization}: the prior $\bm\theta_1\sim\mathcal{N}(\hat{\bm\theta}_1,P_1)$ where $P_1\in\R^{d\times d}$ is positive definite and $\hat{\bm\theta}_1\in\R^d$.

\medskip

{\bf Recursion}: at each time step $t=1,2,\ldots$
\begin{enumerate}
\item Prediction:
\begin{align*}
    & \mathbb{E}\left[y_t\mid (\bm{x}_s,y_s)_{s<t},\bm{x}_t\right] = \hat{\bm\theta}_t^\top f(\bm{x}_t)\,, \\
    & Var\left[y_t\mid (\bm{x}_s,y_s)_{s<t},\bm{x}_t\right] = \sigma^2 + f(\bm{x}_t)^\top P_t f(\bm{x}_t) \,.
\end{align*}
\item Estimation:
\begin{align*}
    & \hat{\bm\theta}_{t+1} = \hat{\bm\theta}_t + \frac{P_t f(\bm{x}_t)}{f(\bm{x}_t)^\top P_t f(\bm{x}_t)+\sigma^2} (y_t-\hat{\bm\theta}_t^\top f(\bm{x}_t)) \,, \\
    & P_{t+1} = P_t - \frac{P_t f(\bm{x}_t)f(\bm{x}_t)^\top P_t}{f(\bm{x}_t)^\top P_t f(\bm{x}_t)+\sigma^2} + Q \,.
\end{align*}
\end{enumerate}
}
\end{algorithm}

There is a wide literature concerning the setting of the hyper-parameters $\hat{\bm\theta}_1,P_1,\sigma^2,Q$ on which the Kalman Filter crucially relies, see for instance~\cite{brockwell2002introduction,durbin2012time,fahrmeir2013multivariate}. We observe that the iterates of $\hat{\bm\theta}_t$ depend only on $\hat{\bm\theta}_1,P_1^*=P_1/\sigma^2$ and $Q^*=Q/\sigma^2$, reducing the set of hyper-parameters as in \cite{brockwell2002introduction}.

An interesting degenerate covariance matrix is the static setting $Q^*=0$ (the state equation becomes $\bm\theta_{t+1}=\bm\theta_t$). Defining $\hat{\bm\theta}_1=0,\ P_1^*=I$, the estimate $\hat{\bm\theta}_t$ is a regularized empirical risk minimizer:
\begin{equation*}
    \hat{\bm\theta}_t = \arg\min_{\bm\theta\in\mathbb{R}^d} \left(\sum\limits_{s=1}^{t-1} (y_s - \bm\theta^\top f(\bm{x}_s))^2 + \|\bm\theta\|^2 \right)\,.
\end{equation*}

In order to obtain a dynamic setting we maximize the likelihood on the training set. The Expectation-Maximization algorithm is a renowned algorithm allowing to find a local optimum. However the lack of global guarantee makes it inefficient in our case, and we chose to apply some kind of grid search. Precisely we decided to set $P_1^*=I$ as in the static setting, and for a given $Q^*$ the optimal $\hat{\bm\theta}_1$ for the likelihood has a closed-form solution. $Q^*$ is of dimension $10\times 10$ and we chose to restrict ourselves to diagonal matrices whose coefficients are in the set $\{2
^j, -30\le j\le 0\}$. This is still a set of around $8\cdot 10^{14}$ elements, thus we used an iterative greedy procedure: we start from $Q^{*(0)}=0$ and at each step, having $Q^{*(k)}$ in hand, we compute the likelihood of each matrix where only one coefficient differ from $Q^{*(k)}$, and we define $Q^{*(k+1)}$ as the one maximizing the likelihood among those tested. This algorithm yielded less than $10^4$ evaluations of the likelihood.

In order to take the lockdown into account in the state-space representation, it is natural to consider a varying state noise covariance $Q_t$. Indeed, we expect the model to change much faster during and after the lockdown than before. It motivates a dynamic estimation of $Q_t$, however due to the amplitude of the crisis we modelled a break in the data at the lockdown beginning. We chose to change only the state noise covariance at the break time $T$, and for $t\neq T$ we use $Q_t^*=0$ in the static setting or $Q_t^*=Q^*$ in the dynamic setting. We don't want to put any {\it a priori} on the break, therefore we defined $Q_T^*=P_1^*=I\gg Q^*$.

\subsection{Correction of the full model}\label{sec:finetuning}

In the previous methods the nonlinear effects $f_j(\cdot)$ were frozen and adjusted with a multiplicative factor. However it may be insufficient on certain new types of behavior. Since learning a new model from scratch is inadvisable considering the few samples of target data available, we would like to start from the previously trained model and adapt it on the few instances available. This is a particular case of the framework of transfer learning, more specifically of model fine-tuning (FT). It consists in reusing a part of the parameters learned on the source set (typically neural network layers) and adjust them with a few gradient iterations on the target one for instance. Model fine-tuning has been successful in different fields such as computer vision \cite{shin2016deep} or even time series forecasting \cite{laptev2018reconstruction}.

In our case we will fine-tune the parameters of our GAM. Since it boils down to a penalized linear regression problem, fine-tuning on it consists in fine-tuning a linear model. This framework was elaborated in \cite{obst2020transfer}. Starting from the coefficients $\hat{\bm\beta}_S$ learned on the source data, for each time step we perform $K$ iterations of batch gradient descent with fixed step size $\alpha$ on following objective function to yield an adjusted parameter vector $\hat{\bm\beta}_t$:

\begin{equation*}
    \mathcal{L}_t(\bm\beta) = \sum_{s=1}^{t-1} \Big(y_s - \sum_{j=1}^d \sum_{k=1}^{m_j} \beta_{j,k} B_{j,k}(x_{s,j}) \Big)^2
\end{equation*}

Let $B(\bm{x}_s)$ be the vector of the $B_{j,k}(x_{s,j})$ and $B(X_t)$ denote the matrix made by the concatenation (by row) of the $B(\bm{x}_s)$ for $s=1, \dots, t-1$. As discussed by the aforementioned paper, the choice of the step size $\alpha$ is not crucial, as long as it is small enough. In practice a good step size is $\alpha=\alpha^* / 5$ where $\alpha^* = 2 / \Big(\lambda_{\max}(B(X_t)^\top B(X_t)) + \lambda_{\min}(B(X_t)^\top B(X_t)) \Big)$ and $\lambda_{\max}(M)$ and $\lambda_{\min}(M)$ respectively designate the maximum and minimum eigenvalue of $M$. Ergo the major hyperparameter to tune is $K$ the number of gradient iterations to perform. Theoretical methods are currently being investigated in the aforementioned paper and have been used to guide our choice here, but it was also observed empirically that for $K$ between 50 and 100 the results are often good. Therefore a number of iterations in that range is always considered, and this choice usually coincides with the suggested theoretical guidelines.

\section{Data and Model Presentation}

In this section we detail the GAM model that has been used to forecast the French electricity consumption, as well as the data on which is has been applied.

\subsection{Presentation of the data}

The French electricity consumption is freely available on the website of the system operator RTE (R\'eseau et Transport d'\'{E}lectricit\'e)\footnote{\url{https://opendata.rte-france.com}}. Our dataset ranges from the 1\textsuperscript{st} of January 2012 to the 7\textsuperscript{th} of June 2020 with a 30 minutes temporal resolution.

As explanatory variables we obtained national averaged temperature on the website of the French weather forecaster M\'et\'eo-France\footnote{\url{https://donneespubliques.meteofrance.fr/}}.
We took observed temperatures instead of forecasts in order to use only open data and make the results reproducible. As our goal is to compare different forecasting strategies on the same data this choice is relevant and allows a more precise comparison as we don't include in the score the uncertainty due to physical meteorological forecast.

We train the models on historical data from the beginning of 2012 to the end of August 2019. In this paper we are interested in predicting the load during and after the COVID-19 lockdown period in France. Since the consumer behavior changed abruptly during the first month and stabilized during the second one, we divide the crisis test data in two periods. The first one ranges from March 16\textsuperscript{th} to April 15\textsuperscript{th} and the second one from April 16\textsuperscript{th} to June 7\textsuperscript{th}. Note that although the lockdown officially begun Tuesday the 17\textsuperscript{th} of March 2020 at midday in France, we consider March 16\textsuperscript{th} as the first day of our lockdown period as the behavior had already changed. Finally, in order to assess the suitability of the offline methods and of the ones that do not model the break we consider the pre-lockdown period between September 1\textsuperscript{st} 2019 and March 15\textsuperscript{th} 2020.


\subsection{The additive model}
The time of day is crucial for load forecasting. It doesn't appear in the following definition of the additive model because we build one model for each instant of day, i.e. we treat the 48 half-hour time series independently:
\begin{align}\label{modele_GAM_FR}
    y_t =\ & \sum\limits_{i=1}^7 \sum\limits_{j=0}^1 \alpha_{i,j}\mathds{1}_{\text{DayType}_t=i}\mathds{1}_{\text{DLS}_t=j} \nonumber\\
    & + \sum\limits_{i=1}^7 \beta_i \text{Load1D}_t \mathds{1}_{\text{DayType}_t=i} + \gamma \text{Load1W}_t \\
    & + f_1(t) + f_2(\text{ToY}_t) + f_3(t, \text{Temp}_t) + f_4(\text{Temp95}_t) \nonumber\\
    & + f_5(\text{Temp99}_t) + f_6(\text{TempMin99}_t, \text{TempMax99}_t) + \varepsilon_t \nonumber\,,
\end{align}
where at each time $t$,
\begin{itemize}
    \item
    $y_t$ is the electricity load for the considered instant,
    \item
    $\text{DayType}_t$ is a categorical variable indicating the type of the day of the week,
    \item
    $\text{DLS}_t$ is a binary variable indicating whether $t$ is in summer hour or winter hour,
    \item
    $\text{ToY}_t$ is the time of year whose value grows linearly from 0 the 1\textsuperscript{st} of January 00h00 to 1 on the 31\textsuperscript{st} of December 23h30,
    \item
    $\text{Temp}_t$ is the temperature,
    \item
    $\text{Temp95}_t$ and $\text{Temp99}_t$ are exponentially smoothed temperatures of smoothing factor $0.95$ and $0.99$,
    \item
    $\text{TempMin99}_t$ and $\text{TempMax99}_t$ are exponentially smoothed variables of factor $0.99$ of the minimal and maximal temperature of the day,
    \item
    $\text{Load1D}$ and $\text{Load1W}$ are the load of the day before and the load of the week before.
\end{itemize}
The models are trained in \texttt{R} using the library \texttt{mgcv} \cite{wood2015package}.

As previously mentioned in Section \ref{section:gam_adapt}, we suppose that $\varepsilon_t$ is a Gaussian noise with 0 mean and constant variance. However this hypothesis is rarely true in practice and we observe an auto-correlation structure in the error. We thus propose to model it with an ARIMA model by selecting the best model with AIC criteria \cite{akaike1978time} in the family of ARIMA(p,d,q) where $p,q\leq 100$ and $d\leq1$ (we use the \texttt{R} function \texttt{auto.arima} of R. Hyndman). In that case the forecast are performed adding GAM forecasts and the short term correction of the ARIMA models exploiting recent observations.

\subsection{Knowledge transfer from Italy}

Italy was the first country to be massively affected by the novel coronavirus in Europe. The Italian government decreed a total lockdown from the 9\textsuperscript{th} of March 2020, hence 7 days before the French one. Also it seems reasonable to make the assumption that countries will respond to the same stay-at-home order in similar ways, which is reasonable considering Figure \ref{fig:evolution_covid}. Hence our idea is to use this one week head-start and to adjust our GAM model for France accordingly to the changes observed in Italy. We have at our disposal data from the Italian system operator Terna\footnote{\url{https://www.terna.it}} and meteorological data gathered through the \texttt{R} package \texttt{Riem} available from the 1\textsuperscript{st} of January 2015 to the 28\textsuperscript{th} of June 2020 with a 1 hour temporal resolution. For each instant, a model similar to (\ref{modele_GAM_FR}) is constructed on the data on the range 2015-2019, with the main differences being that the effects $f_3(\cdot)$ and $f_6(\cdot)$ are removed, and that $f_2(\cdot)$ is replaced by a sum of 7 effects, one for each day of the week. Then the same procedure as described in Section~\ref{sec:finetuning} is applied. Let $\hat{\bm\delta}_t$ denote the adjustment of the estimated coefficients obtained by performing the aforementioned fine-tuning procedure on the Italian data ranging from the 28\textsuperscript{th} of February up to date $t$ (typically $t$ could correspond to the 15\textsuperscript{th} of March, the day before the stay-at-home order begun in France). We then use $\Tilde{\bm\beta}_t = \hat{\bm\beta}_S^{FR} + \rho \, \hat{\bm\delta}_t$ to perform the predictions for France, where $\hat{\bm\beta}_S^{FR}$ is the French source parameters vector and $\rho$ is a scale parameter accounting for the difference of load levels between France and Italy. We refer to this model as GAM-$\delta$. Since the ToY effect is modelized differently for the Italian model (one function per day of the week), we will not adjust the corresponding coefficients in the French model. This is further justified by the fact that in general the ToY effect is very specific to a country, and it should be learned on a whole year at least. As for the choice of $\rho$, making the assumption that the consumption in France and Italy are proportional with a factor $\rho$ allows us to use the simple estimate $\hat{\rho} = \sum_{t} y_t^{FR} / \sum_{t} y_t^{IT}$ summed over a year for instance. The advantage of GAM-$\delta$ is that it can be applied to reduce the prediction error starting at the very first day of lockdown. One can afterwards combine this procedure with fine-tuning on the eventually available French data. The procedures for both regular fine-tuning and GAM-$\delta$ are summarized in Algorithm \ref{alg:gam_delta}.

\begin{algorithm}[t]
{\caption{Transfer learning at time step $t$}
\label{alg:gam_delta}}
{
{\bf Inputs}: Step size $\alpha$, number of iterations $K$, French and Italian source parameters $\hat{\bm\beta}_S^{FR},\hat{\bm\beta}_S^{IT}$, scale parameter $\rho$.

\medskip

{\bf If GAM fine-tuned:}
\begin{itemize}
    \item
    Initialize $\hat{\bm\beta}_t \leftarrow \hat{\bm\beta}_S^{FR}$.

    \item Repeat $K$ times:\\
    \hspace{1cm}$\hat{\bm\beta}_{t} \leftarrow \hat{\bm\beta}_{t} -\alpha \nabla \mathcal{L}_{t-1}^{FR} (\hat{\bm\beta}_{t})$.
        
    \item Predict $\hat{y}_t = \hat{\bm\beta}_t^\top B(\bm{x}_t)$.
\end{itemize}

\medskip

{\bf If GAM-$\delta$:}
\begin{itemize}
    \item
    Initialize $\hat{\bm\beta}_t^{IT} \leftarrow \hat{\bm\beta}_S^{IT}$.

    \item Repeat $K$ times:\\
    \hspace{1cm}$\hat{\bm\beta}_{t}^{IT} \leftarrow \hat{\bm\beta}_{t}^{IT} -\alpha \nabla \mathcal{L}_{t-1}^{IT} (\hat{\bm\beta}_{t}^{IT})$.
    
    \item Set $\hat{\bm\delta}_t = \hat{\bm\beta}_{t}^{IT} - \hat{\bm\beta}_S^{IT}$, $\tilde{\bm\beta}_t = \hat{\bm\beta}_S^{FR} + \rho \, \hat{\bm\delta}_t$.
    
    \item Predict $\hat{y}_t = \tilde{\bm\beta}_t^\top B(\bm{x}_t)$.
\end{itemize}

\medskip

{\bf If GAM-$\delta$ fine-tuned:}
\begin{itemize}
    \item Do fine-tuning on Italian data: $\tilde{\bm\beta}_t = \hat{\bm\beta}_S^{FR}+\rho \, \hat{\bm\delta}_t$.
    \item Repeat $K$ times:\\
    \hspace{1cm}$\tilde{\bm\beta}_t \leftarrow \tilde{\bm\beta}_t -\alpha \nabla \mathcal{L}_{t-1}^{FR} (\tilde{\bm\beta}_t)$.
    \item Predict $\hat{y}_t = \tilde{\bm\beta}_t^\top B(\bm{x}_t)$.
\end{itemize}
}
\end{algorithm}

\section{Experiments}

The presented adaption methods are used for the French electricity load forecasting problem. While accuracy metrics are of paramount importance, we also focus on the interpretation of our results and on model behavior.

\subsection{Model dynamics}

\begin{figure}[t]
    \centering
    \includegraphics[scale=0.4]{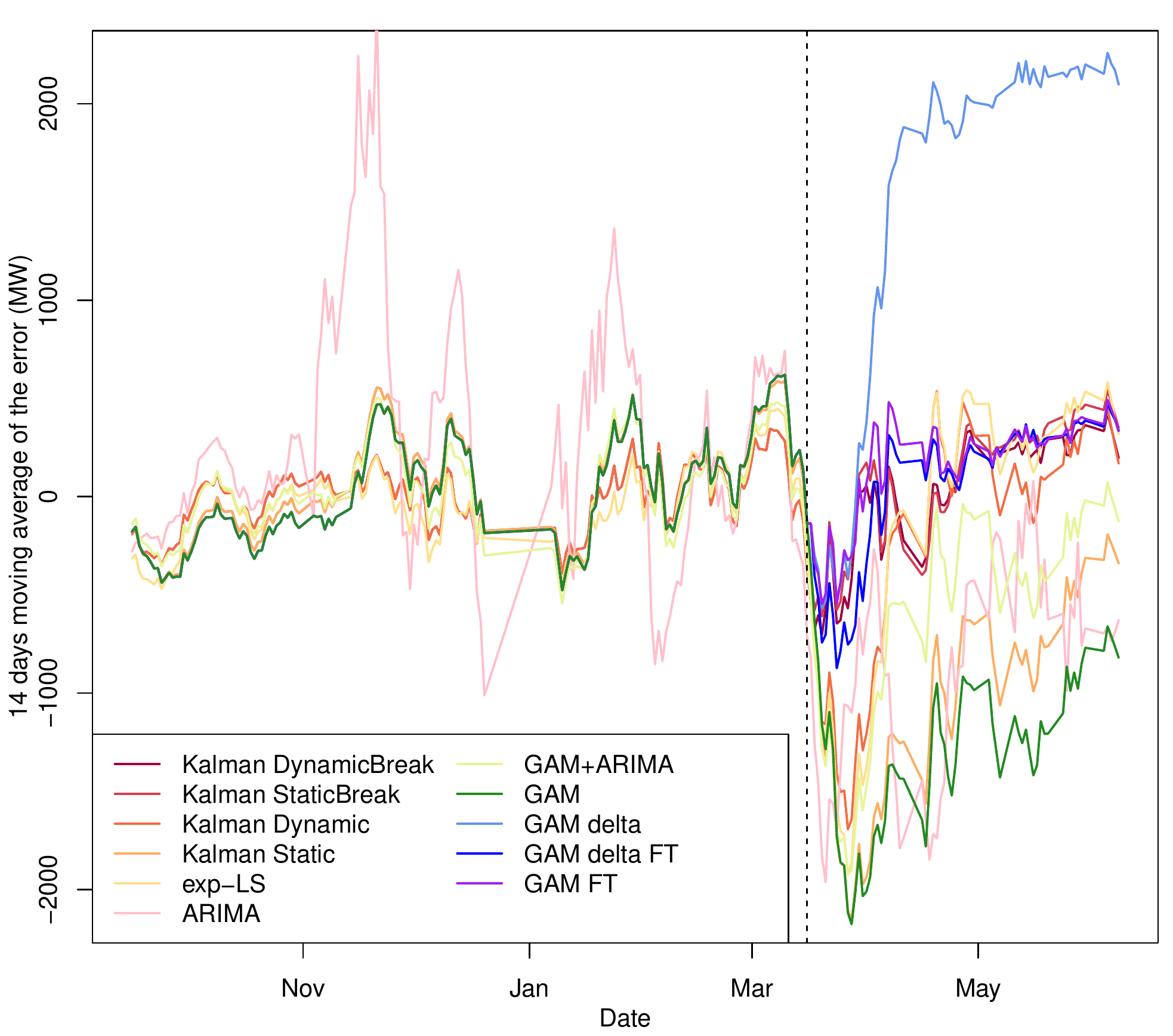}
    \caption{Moving average of the error of the different models at 8-8:30 PM.}
    \label{fig:error_ma}
\end{figure}

The moving average of the error of the different models are represented in Figure \ref{fig:error_ma}. At the beginning of the lockdown all the models will tend to overpredict the load. However most of our adaptive methods quickly accommodate to the lower demand and progressively reduce their bias, notably Kalman with dynamic break and GAM fine-tuned. On the contrary regular GAM does not succeed in reducing the error (even with the help of an ARIMA) as it keeps overpredicting the demand. GAM-$\delta$ on the other hand is very good during the first couple of days, efficiently taking advantage of the change in patterns observed in Italy. However it quickly drifts away over time because the Italian consumption recovers faster than the French one during the second month of lockdown (see fig. \ref{fig:evolution_covid}). However since the objective of GAM-$\delta$ is to provide an initial boost of performance during the first couple of weeks while the other models adjust, this is only a minor issue (see Section \ref{sec:aggregation}).

\begin{figure*}[t]
\centering
\subfloat[Static]{\includegraphics[width=2.2in]{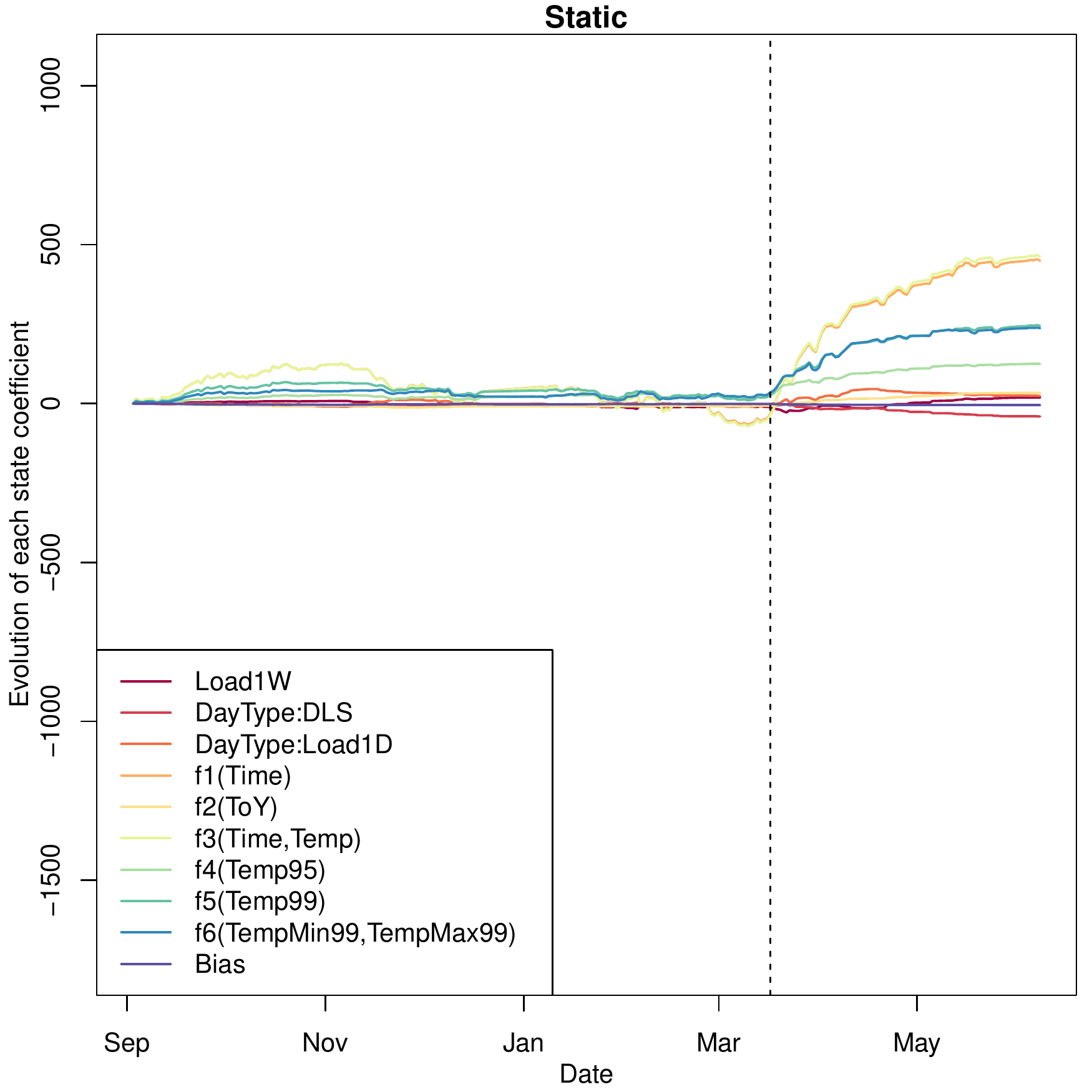}
\label{lab}}
\hfil
\subfloat[Dynamic]{\includegraphics[width=2.2in]{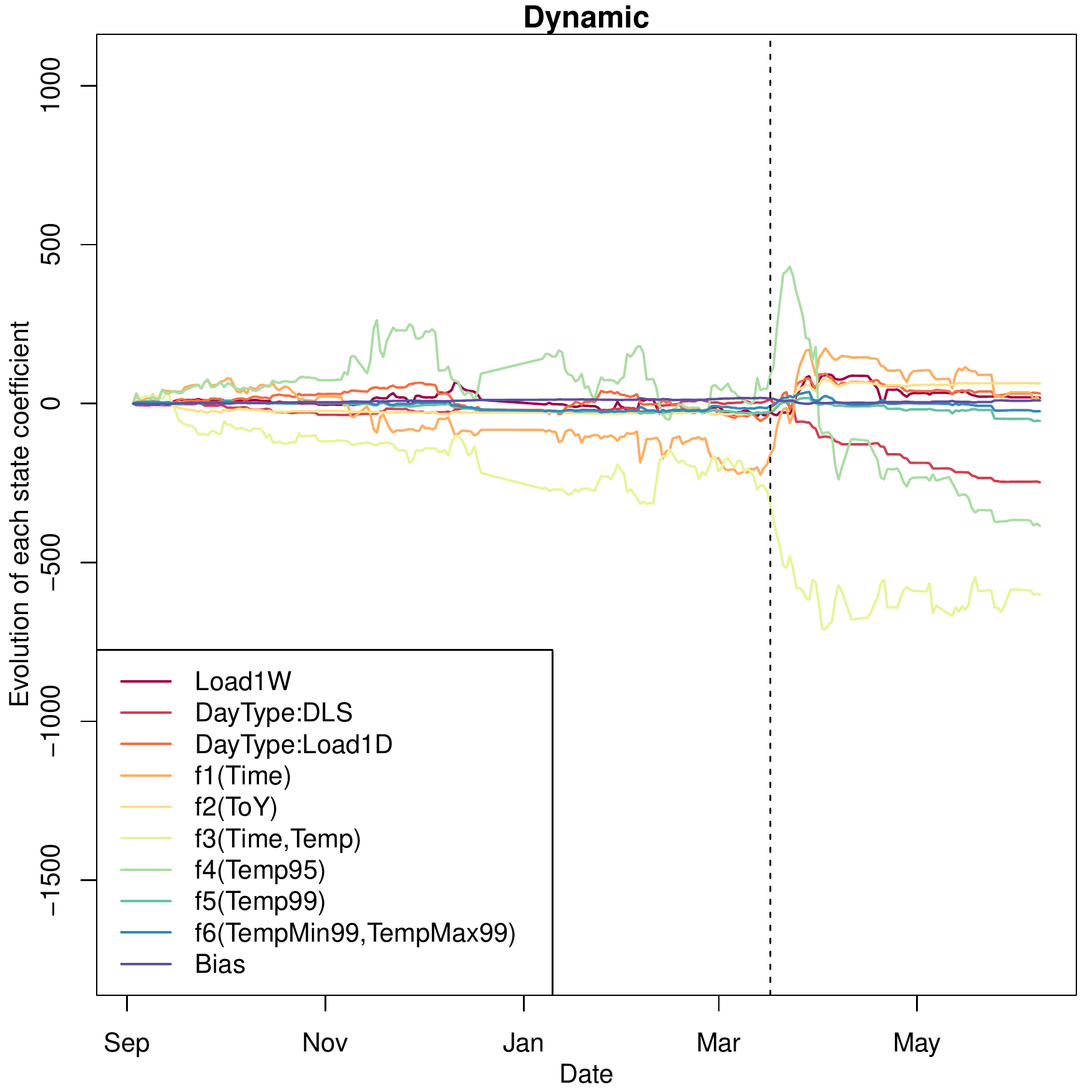}%
\label{lab}}
\hfil
\subfloat[Dynamic with break]{\includegraphics[width=2.2in]{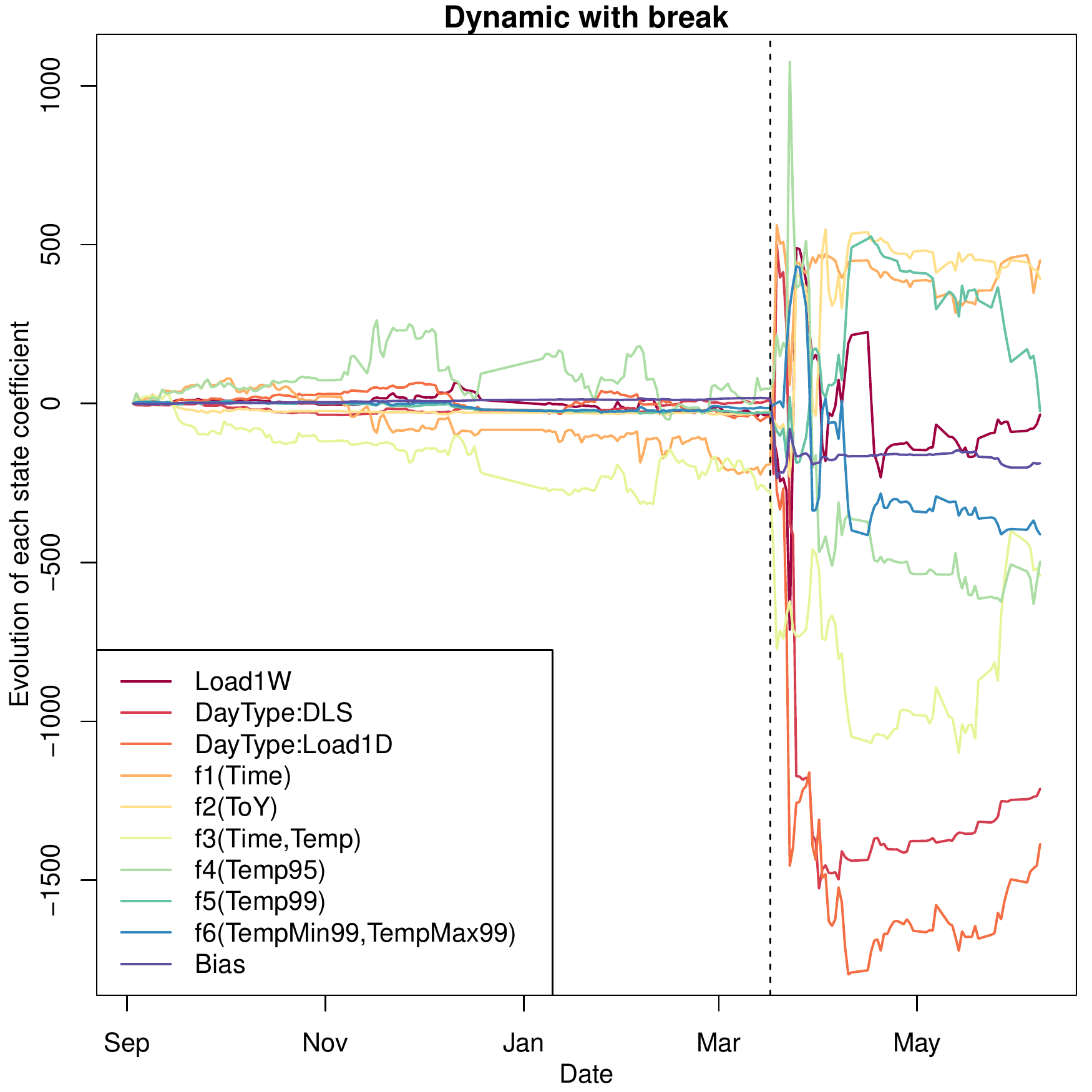}%
\label{lab}}
\caption{Evolution of the state coefficients for different Kalman variants at 8-8:30 PM (subtracting the coefficients on September 1\textsuperscript{st} 2019).}
\label{fig:kalman_evol}
\end{figure*}

We test the Kalman filter in a static and a dynamic setting as described in Section \ref{section:kalmanadaptation}. For both we assess the introduction of a break at lockdown.
The evolution of the state estimate $\hat{\bm\theta}_t$ is displayed in Figure \ref{fig:kalman_evol} for different settings.
In the static setting the Kalman filter optimizes a state which is assumed to be constant, hence explaining a slow evolution compared to the faster changes of the dynamic one. Moreover, the model changes faster during lockdown in both settings. As expected the introduction of a break covariance matrix at the beginning of the lockdown allows the model to adapt much faster.

\begin{figure}[h]
    \centering
    \includegraphics[scale=0.4]{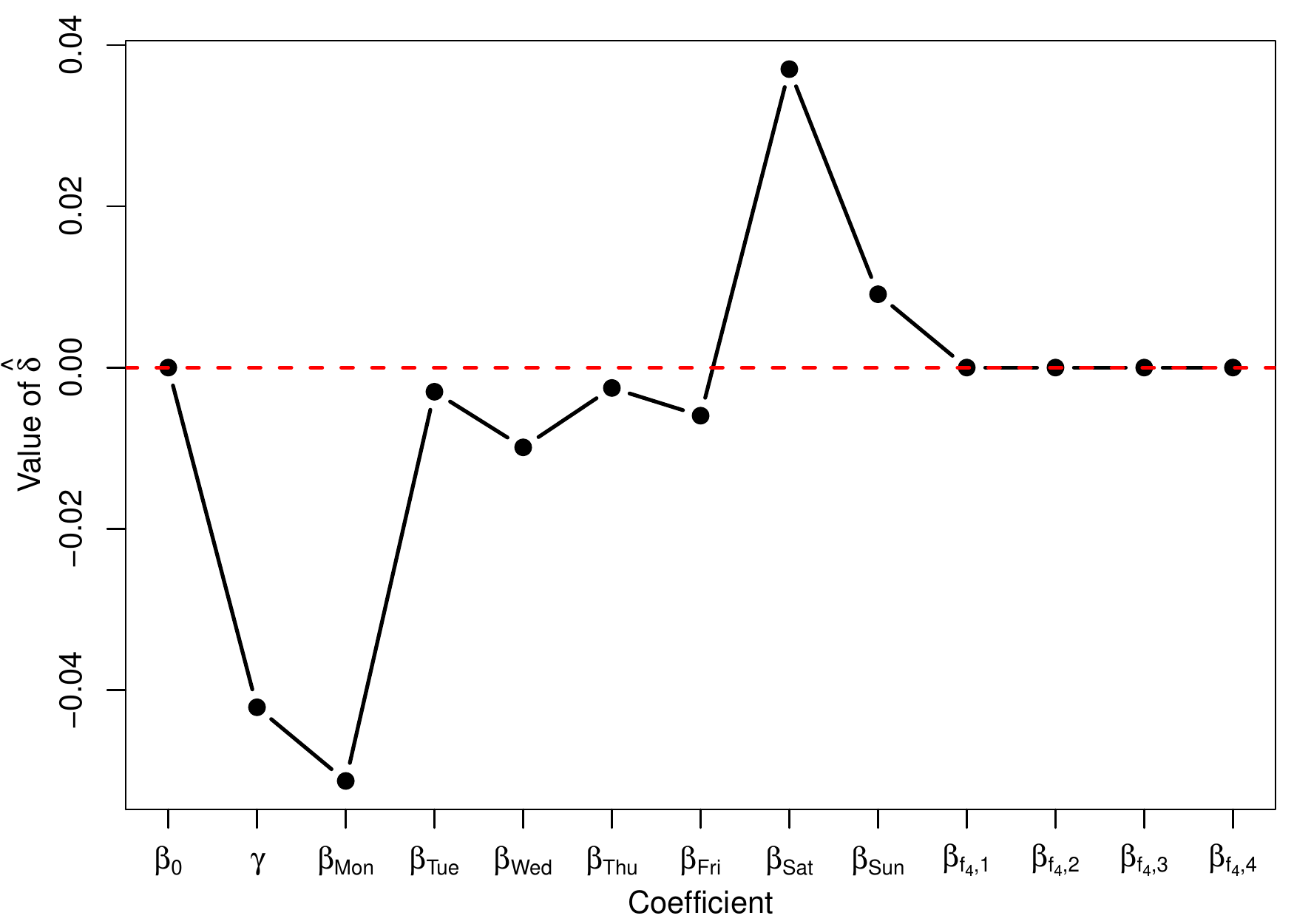}
    \caption{Value of $\hat{\bm\delta}_t$ fine-tuned on the period 16/03-15/04 at at 8-8:30 PM.}
    \label{fig:evo_delta}
\end{figure}

The model dynamics can be analysed for the fine-tuning too. The only coefficients of $\hat{\bm\delta}_t$ with a significant evolution after fine-tuning are the ones pertaining to the lagged load ($\gamma$ for Load1W and $\beta_i, i=1..7$ for Load1D) and have been represented in Figure \ref{fig:evo_delta}. The other ones are zero and have been omitted for clarity. The coefficients of the working days drop, especially the Monday, whereas the ones of the weekend increase, notably Saturday. It can be interpreted as follows: the historical model learned a certain transition between the different days of the week. With the lockdown now all the days are similar and close to a Saturday, which has a lower demand than Monday and thus the associated coefficient plummets. The coefficient of Saturday soars because the demand on Fridays is now much lower than it used to be and that daily profiles are similar. Finally since during the first weeks the electricity demand progressively decreases (see Fig. \ref{fig:evolution_covid}) the coefficient of $\gamma$ drops as well.

\subsection{Aggregation}\label{sec:aggregation}

We proposed 2 load forecasting models (ARIMA, GAM) and different variants to adapt them to the lockdown period (exp-LS, Kalman adaptation, transfer learning) leading to 11 candidates. A natural approach is then to aggregate them in a single forecast which will take benefit of the best one in function of time. This is the main idea behind online aggregation methods which has already demonstrated its benefits in the field of electricity load forecasting (see \cite{gaillard2015forecasting}, \cite{goehry2019aggregation}). Since Figure~\ref{fig:profile_covid} shows the convergence of the daily profiles towards the Saturday shape, this as well as \cite{nagbe2020covid} motivates adding another expert named GAM Saturday, where the prediction is made by the regular GAM as if every day was a Saturday.

We recall briefly the main principles of the online aggregation approach and refer the interested reader to \cite{Cesa-Bianchi:2006} for a complete presentation. A bounded sequence of observations (here half-hourly total consumption of customers) $y_1,\dots,y_n \in [0,B]$ is observed ($B$ being an unknown constant). We have access to a set of $N$ experts who produce forecasts of the sequence at each instant $t$ based on past values of $y$. After that, aggregation is computed step by step:  $\hat y_t =  \sum_{j=1}^N \hat{p}_{j, t} \hat{y}_{t}^j$ where the weights are updated according to past performances of each experts. To compute the weights we use the ML-Poly algorithm of \cite{gaillard2014second}, implemented in the \texttt{R} package \texttt{OPERA} \cite{gaillard2016opera}. To summerize the procedure, the algorithm puts more weight on an expert which improves the performance of the aggregation in the past using a gradient descent like strategy with a vectorial time varying step (also called the learning rate) $\eta_{k,t}$ depending only on the past performances of the experts so that no parameter tuning is needed. Finally a few experts are introduced in the aggregation only at lockdown. Indeed the transfer learning experts don't make sense (there is no target data), the Kalman experts modelling the break coincide with the other ones before lockdown, and the expert considering every day is a Saturday was only introduced for the lockdown period. These specialized experts are added to the aggregation at the lockdown period with a uniform weight ($1/12$), and the experts present before share the rest of the weight proportionally to their previous weight \cite{devaine2013forecasting}.

\begin{figure}[t]
    \centering
    \includegraphics[scale=0.46]{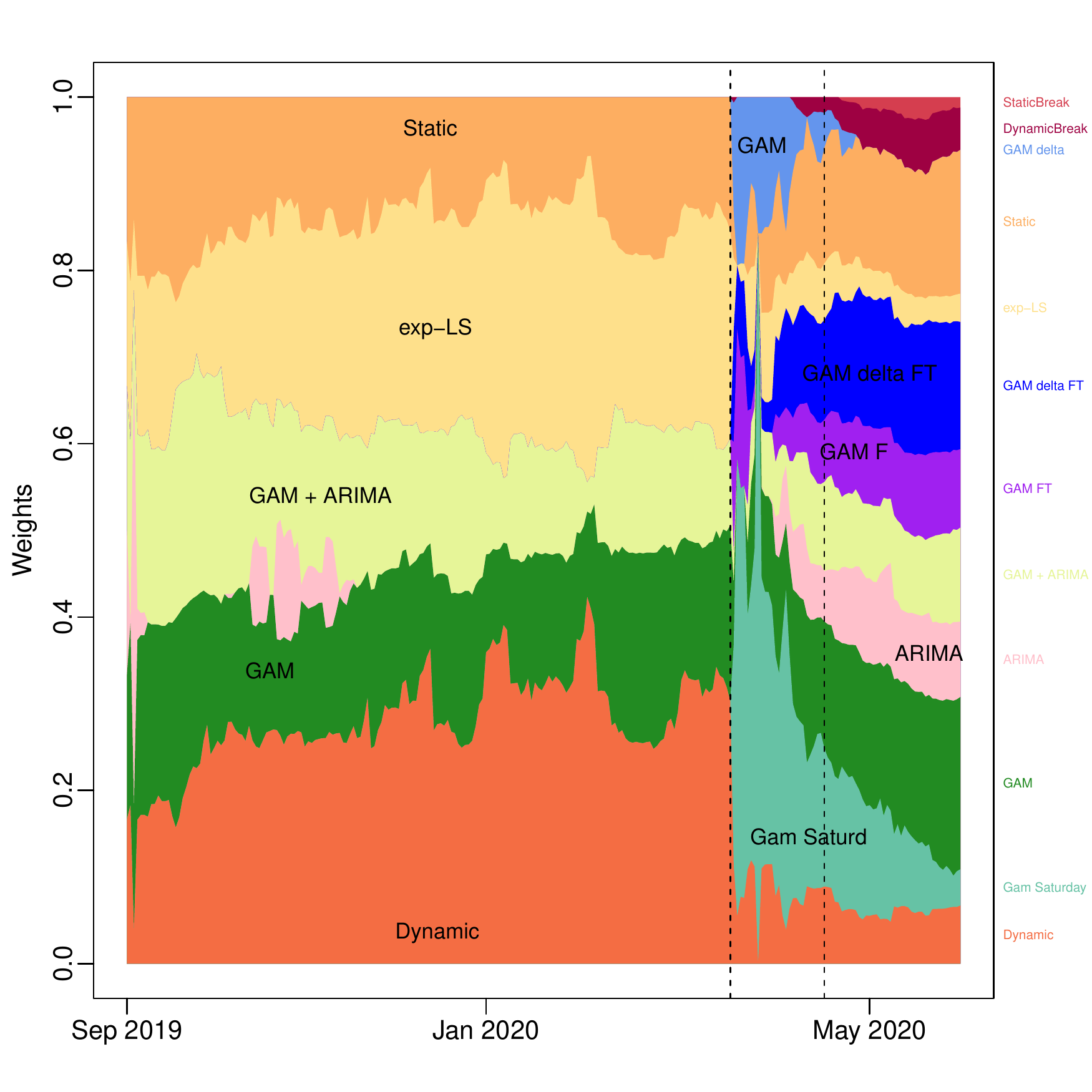}
    \caption{Weights attributed to each expert by the aggregation method at 8-8:30 PM. Dashed lines split the test sets.}
    \label{fig:weights_opera}
\end{figure}

The evolution of the weights of the experts over time is given in Figure \ref{fig:weights_opera}. It gives insight on which predictions are the most useful in the aggregation at a given time. The lockdown acts as a break and causes a significant shift in the weights distribution. As such, GAM Saturday immediately takes a large weight: this is due to the aforementioned resemblance between the daily profiles during the lockdown with Saturdays. Moreover, this expert predicts a lower consumption than reality, compensating for the overestimation of the other experts at the beginning of the lockdown. GAM-$\delta$ also has high importance, as it has knowledge of what happened in Italy and thus suits the new patterns of load demand in France. For instance on the two first days of lockdown (16 and 17\textsuperscript{th} of March) GAM-$\delta$ yields 1984 MW of RMSE, compared to 2674 and 3005 for Dynamic Break and regular GAM respectively. However their importance dwindle with time as the adaptive Kalman and fine-tuning methods have seen enough data and have become more competitive.

\subsection{Numerical results}

As usual in electricity load forecasting, the performance metrics is the root mean squared error  (in MW) and the mean absolute percentage error (in \%):
\begin{align*}
    \text{RMSE} = \sqrt{\dfrac{1}{n} \sum_{t=1}^n \big(y_t - \hat{y}_t\big)^2}\,, \quad
    \text{MAPE} = \dfrac{100}{n} \sum_{t=1}^n \big|\dfrac{y_t-\hat{y}_t}{y_t}\big| \,,
\end{align*}
where $n$ is the number of instances in the test set.

We display the numerical performance of our methods in Table~\ref{tab:numerical_results}. The benefits brought by any of our methods is clear, with RMSE and MAPE that are significantly lower than a standard GAM+ARIMA on both COVID-19 test sets. The Kalman with Dynamic Break yields the best results for the two error metrics on both test sets, but the fine-tuned methods are very close to it. The additional benefits brought by expert aggregation is emphasized by the two last rows. The algorithm manages to take advantage of the individual specificities of the different predictions, leading to further error reduction on both test periods. It is interesting to note that while individually poor (see Table \ref{tab:numerical_results}), the inclusion of the GAM Saturday in the mixture is of paramount importance for the first testing period. This is because it compensates for the bias of the other experts (they tend to overestimate the demand whereas GAM Saturday underestimates it).

\begin{table*}[t]
    \caption{Numerical performance in MAPE (\%) and RMSE (MW).}
    \label{tab:numerical_results}
    \centering
    \begin{tabular}{|c|c|c|c|}
        \hline
        Method & 2019/09/01 - 2020/03/15 & 2020/03/16 - 2020/04/15 & 2020/04/16 - 2020/06/07 \\
        \hline\hline
        ARIMA & 4.10 \%, 3341 MW & 5.44 \%, 3248 MW & 5.59 \%, 3135 MW \\
        \hline\hline
        GAM & 1.39 \%, 1085 MW & 4.83 \%, 2961 MW & 3.12 \%, 1753 MW \\
        \hline
        GAM + ARIMA & 1.34 \%, 1050 MW & 4.28 \%, 2654 MW & 2.65 \%, 1464 MW \\
        \hline
        \hline
        exp-LS & 1.26 \%, 982 MW & 3.94 \%, 2521 MW & 1.97 \%, 1029 MW \\
        \hline
        \hline
        Kalman Static & 1.38 \%, 1077 MW & 4.81 \%, 2923 MW & 2.85 \%, 1588 MW \\
        \hline
        Kalman StaticBreak & - & 2.79 \%, 1954 MW & 1.59 \%, 855 MW \\
        \hline
        Kalman Dynamic & {\bf 1.26 \%, 979 MW} & 3.66 \%, 2351 MW & 1.89 \%, 1002 MW \\
        \hline
        Kalman DynamicBreak & - & {\bf 2.73 \%, 1902 MW} & {\bf 1.62 \%, 854 MW} \\
        \hline\hline
        Fine-tuned & - & 2.78 \%, 1917 MW & 1.80 \%, 938 MW \\
        \hline
        GAM $\delta$ & - & 4.11 \%, 2364 MW & 6.09 \%, 2713 MW \\
        \hline
        GAM $\delta$ - Fine-tuned & - & 2.81\%, 1912 MW & 1.72 \%, 905 MW \\
        \hline
        \hline
        GAM Saturday & 8.33 \%, 6425 MW & 6.09 \%, 3970 MW & 8.40 \%, 4616 MW \\
        \hline
        \hline
        Aggregation without GAM Saturday & 1.28 \%, 1005 MW & 3.01 \%, 2014 MW & {\bf 1.44 \%, 745 MW} \\
        \hline
        Aggregation with GAM Saturday & 1.28 \%, 1005 MW & {\bf 2.54 \%, 1636 MW} & {\bf 1.49 \%, 766 MW} \\
        \hline
    \end{tabular}
\end{table*}

\section{Conclusion}

In this paper, we proposed two novel approaches of adaptive generalized additive models, one relying on Kalman filtering and the other on transfer learning with GAM fine-tuning. Kalman philosophy consists in reacting quickly to a change in the data and update the forecasting taking advantage on recent observations. Transfer allows to share information from other data sets with similar/complementary properties. The methods have been applied on real French electricity consumption data from the COVID-19 lockdown period. We show the benefits of the transfer approach to anticipate the lockdown effect using Italian data and demonstrate the efficiency of adaptive methods to significantly improved predictions compared to benchmark models without relying on the inclusion of new exogenous features. Moreover expert aggregation enabled to take advantage of the individual experts' specificities and enhanced the results even further.

While in this paper we focused on adapting GAM, the proposed framework can be applied to other approaches. The use of neural networks for instance, with their high performance in the field of load forecasting, will soon be investigated.

We also plan to include other exogenous information as mobility data proposed in \cite{chen2020using} or macro-economic indicators. Regarding load data, we believe that exploiting regional data could be pertinent as the propagation of the pandemic and its impact on consumption was different depending the region in France and Italy. Also, we would like to include more countries. For these next steps, transfer approaches will obviously be of fundamental importance but also adaptive questions  as the effect of this exogenous variables will probably vary with time.


%

\appendices


\ifCLASSOPTIONcaptionsoff
  \newpage
\fi



\bibliographystyle{IEEEtran}
\bibliography{IEEEabrv,biblio.bib}
\end{document}